# A Structured Methodology For Spreadsheet Modelling

Brian Knight, David Chadwick, Kamalesen Rajalingham
University of Greenwich, Information Integrity Research Centre,
School of Computing and Mathematics, 30 Park Row, London SE10 9LS UK
{b.knight, d.r.chadwick, k.rajalingham} @greenwich.ac.uk,

**ABSTRACT**

*In this paper, we discuss the problem of the software engineering of a class of business spreadsheet models. A methodology for structured software development is proposed, which is based on structured analysis of data, represented as Jackson diagrams. It is shown that this analysis allows a straightforward modularisation, and that individual modules may be represented with indentation in the block-structured form of structured programs. The benefits of structured format are discussed, in terms of comprehensibility, ease of maintenance, and reduction in errors. The capability of the methodology to provide a modular overview in the model is described, and examples are given. The potential for a reverse-engineering tool, to transform existing spreadsheet models is discussed.*

## 1. INTRODUCTION

This paper describes an outcome from research done by the authors at the Information Integrity Research Centre at Greenwich over the past 3 years, concerning the problems of the quality of spreadsheet models. The research has focused on the class of business models, including functional formulae, referencing and replication of individual cells and ranges. Surveys have shown that the frequency and severity of errors in spreadsheets is now reaching dangerous proportions. A **KPMG**[9] survey of financial models based on spreadsheets found that 95% of models were found to contain major errors (errors that could affect decisions based on the results of the model), 59% of models were judged to have 'poor' model design, 92% of those that dealt with tax issues had significant tax errors and 75% had significant accounting errors.

There is much evidence[10] that these errors are caused by untrained or badly trained modellers and, that even those who are technically capable of developing applications have not been trained in any development methodology Development is in many ways comparable to the days for main-line software development before the advances due to structured programming and design.

The approach of this research has been to examine the applicability of main-line software-engineering techniques to the very special needs of spreadsheet developers. These needs are partly determined by the visual nature of spreadsheets and their heavy reliance on referencing and intermediate data, and partly by the likely acceptance of techniques within the industry. However sound a methodology is, we cannot expect modellers to undergo much training in software engineering. Object orientation may be technically ideal, but not if modellers have to learn the Unified Modelling Language first.

The aim of the research was to create a methodology for spreadsheets which improved the quality of models, whilst not imposing an extra burden of modellers. To this end, we have looked for a support tool to assist in spreadsheet structuring. Ideally, the tool should be able to take existing models, and transform them to the appropriate form.

Several structured programming and design methodologies originated during the 60s and 70s, with goals to systematise the process of analysis and design of software. The goals were to increase productivity, reduce errors, ease problems of maintenance, and where possible to automate the development process. Amongst these, several important "data-oriented" methods were proposed, amongst which were the Warnier-Orr[1,2] methodology, M.A. Jackson's JSD[3] and Chen's E-R data modelling[4]. These methodologies concentrate primarily on the logical structure of the data, which is likely to be more stable than the software functions. It is argued that this provides a good basis for comprehensible software, which is able to support change and maintenance over time.

In this research, the suitability of a methodology based on Jackson charts for spreadsheet modelling has been investigated. It appears that there are several possible advantages to the adoption of a structured method based on a Jackson data oriented approach. These advantages are may be summarised as:

- A clear modularisation principle,
- A top-level overview of module structure,
- A structured 'indented' format to the layout of module,
- The possibility of automatic structuring of existing spreadsheets.

In section 2 of this paper, we show explain the methodology with illustrations. In section 3 we explain the modularisation principle and the relation to Jackson charts. In section 4 the possibility of automatic re-engineering of existing spreadsheets is discussed.

## 2. APPLICATION OF JACKSON CHARTS TO A SINGLE MODULE

The essence of JSD is the structure diagram and its relationship to block structure, with its three key constructs of sequence, repetition and selection. Figure 1 shows a structure diagram, representing a typical block structured module. Here asterisked blocks are repeated, and blocks marked with an O are selections (mutually exclusive). The diagram shows that A consists of a repeated block B, and each B is made up of either C or D. C is a sequence of block E followed by block F.

The indented structure on the right of figure 1 is the structured programming equivalent of the structure diagram. The philosophy of structured programming, as outlined in [5] promotes the indented form for code. This form has led to huge improvements in the comprehension of code, leading to improvements in productivity, auditing and maintenance. Later work [6] proposed methods for the translation of data structure into structured form. Jackson proposed that the form of the data structure diagram should be extracted from the natural structure existing in the data to be processed.

Some of these techniques can in fact be transferred to the production of spreadsheets, and that this can give a more comprehensible format for spreadsheets, based on indented format. The derivation of the structure charts can be based on the natural data dependencies within the spreadsheet. This is an analytical exercise which depends on a close examination of the semantics of the data involved, to build a logical model in chart form. However, it will be noticed that structure diagrams bear resemblance to the graphs obtained using auditing tools on existing spreadsheets. This reflects the fact that the logical structure is in fact embedded in existing spreadsheets, and may be extracted from them automatically.

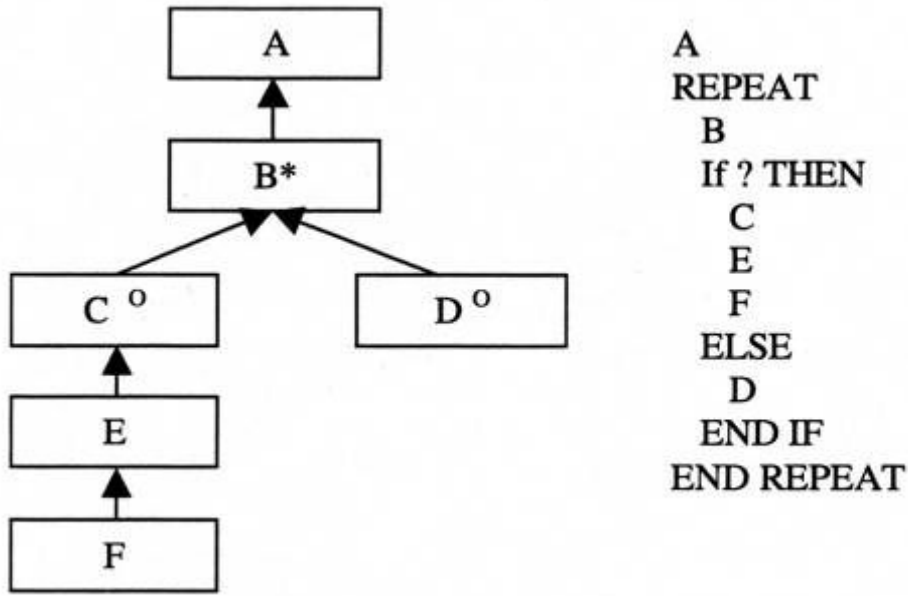

**Figure 1 An example structure diagram**

We first illustrate how these principles can be used to structure a single spreadsheet, leaving a discussion of module formation to the next section. We take as illustration the example of a profit and loss account' l and shown in Figure 2 below. From knowledge of the meaning of the data, we may construct the chart shown in Figure 3.

```
T Howe Ltd.
Trading and Profit and Loss Account for the year
ended 31 December 19X4 (based on Wood, 1996 p. 608)

                                              £          £
Sales                                                135,486
Less Cost of goods sold
   Opening stock                          40,360
   Add Purchases                          72,360
   Add Carriage inwards                    1,570
                                         114,290
   Less Closing stock                     52,360    61,930
Gross Profit                                         73,556
Less Expenses
   Salaries                               18,310
   Rates and occupancy                     4,515
   Carriage outwards                       1,390
   Office expenses                         3,212
   Sundry expenses                         1,896
   Depreciation: Buildings                 5,000
   Depreciation: Equipment                 9,000
   Directors' remuneration                 9,500    52,823
Net profit                                           20,733
Add Unappropriated profits from last year            15,286
                                                     36,019
Less Appropriations
   Proposed dividend                      10,000
   General reserve                         1,000
   Foreign exchange                          800    11,800
Unappropriated profits carried to next year          24,219
```

**Figure 2 An example unstructured spreadsheet**

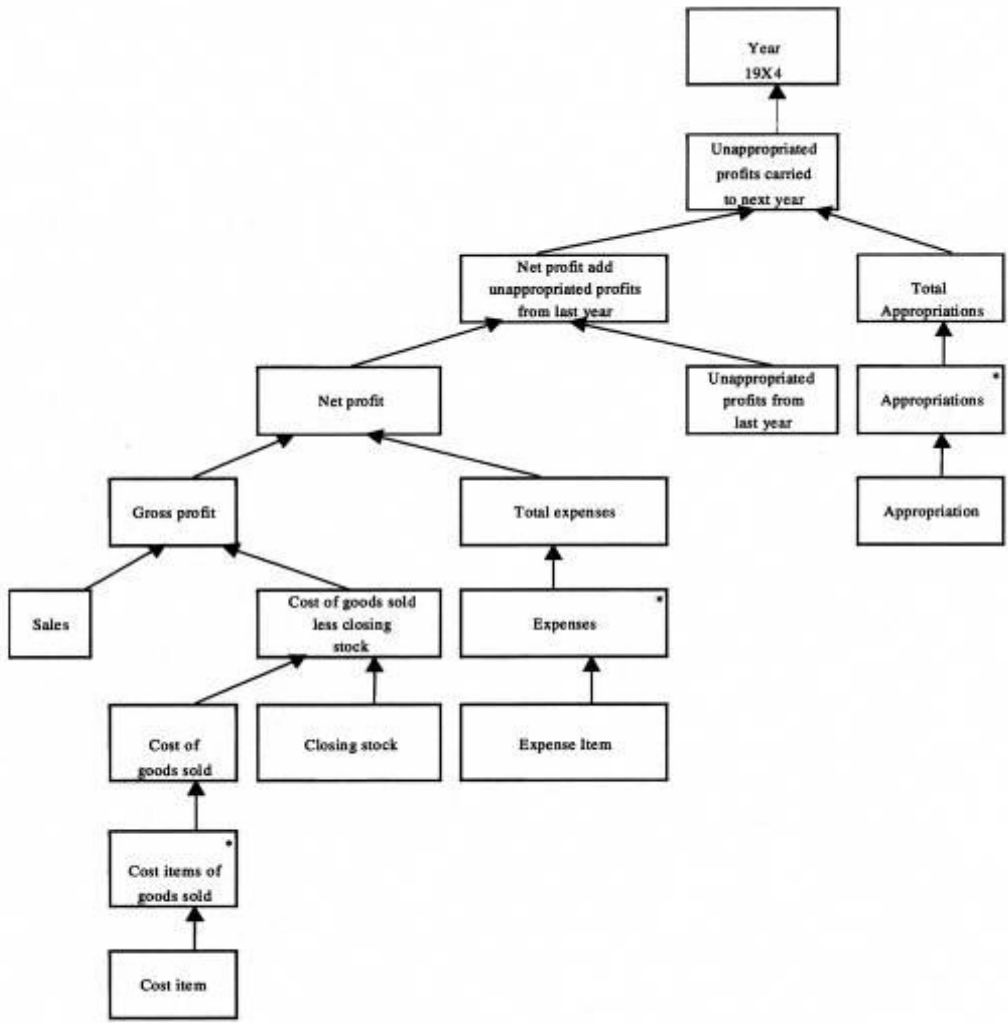

Figure 3: Structure Chart for the Profit and Loss example

To maintain this structure in the spreadsheet view, we can use the indentation principle both on the row labels and on the data values themselves. In fact, we can also insist that data values are indented by assigning a spreadsheet column to each level of indentation. If this is done, the spreadsheet takes on the form shown in Figure 4. Notice that both the semantics and the data are clarified in this layout. For example, we can see straight away on the semantic level *that Unappropriated profits carried to next year is* derived from two figures: *Net Profit add unappropriated profits* from last year and *Total appropriations*. On the data level we see that 24,219 is made up from 36,019 and 11,800. Likewise, we see immediately the constituents of *Total expenses* are a total of eight different expense types, and the data level. Notice also that columns in the spreadsheet show figures on the same semantic level, enabling valid comparisons between figures to be made. For example, column 3 shows *net profit, unappropriated profits from last year, proposed dividend, general reserve, and foreign exchange*. These figures give a valid impression of the state of the trading account at this level of detail. If we were to include a figure from a different level, e.g. *purchases* (from column 7), it would confuse the picture, since it has already been included in net profit.

| | £ | £ | £ | £ | £ | £ | £ |
|---|---|---|---|---|---|---|---|
| Unappropriated profits carried to next year | 24,219 | | | | | | |
| Net Profit *add* unappropriated profits from last year | | 36,019 | | | | | |
| Net profit | | | 20,733 | | | | |
| Gross Profit | | | | 73,556 | | | |
| Sales | | | | | 135,486 | | |
| Cost of goods sold *less* closing stock | | | | | 61,930 | | |
| Cost of goods sold | | | | | | 114,290 | |
| Opening stock | | | | | | | 40,360 |
| Purchases | | | | | | | 72,360 |
| Carriage inwards | | | | | | | 1,570 |
| Closing stock | | | | | | 52,360 | |
| Total expenses | | | | | 52,823 | | |
| Salaries | | | | | | 18,310 | |
| Rates and occupancy | | | | | | 4,515 | |
| Carriage outwards | | | | | | 1,390 | |
| Office expenses | | | | | | 3,212 | |
| Sundry expenses | | | | | | 1,896 | |
| Depreciation: Buildings | | | | | | 5,000 | |
| Depreciation: Equipment | | | | | | 9,000 | |
| Directors' remuneration | | | | | | 9,500 | |
| Unappropriated profits from last year | | | 15,286 | | | | |
| Total appropriations | | 11,800 | | | | | |
| Proposed dividend | | | 10,000 | | | | |
| General reserve | | | 1,000 | | | | |
| Foreign exchange | | | 800 | | | | |

**Figure 4 A structured spreadsheet form**

## 3. MODULARISATION OF SPREADSHEETS

Modularisation is the key to successful software engineering, allowing complex systems to be broken down into manageable sub-systems, for ease of comprehension and maintenance. Indeed, the basic principle guiding modularisation can be said to characterise different software engineering methodologies. Object-oriented software engineering is characterised by Parnas's information hiding principle[7], and Stevens, Constantine and Myers' structured approach[8] is characterised by the concept of code cohesion. In the spreadsheet methodology described here, modules are defined by graphical properties of data structure diagrams.

In section 2, we looked at a structure diagram which took the form of a tree, and showed how this could lead to a structured spreadsheet form for a single module. However, not all spreadsheets are of this simple form, but have structure charts in the form of a more general graph. shows an example of such a chart. The chart is different to that in Figure 1 in that there is a loop in the relationships connecting A B and Q so that we do not any longer have a tree form. In this chart, data block C contributes to block A and to block B. We can of course turn the graph into a tree by duplicating the structure C, D,E, as shown in figure 5. However, the resulting structured spreadsheet will then have to include the rows

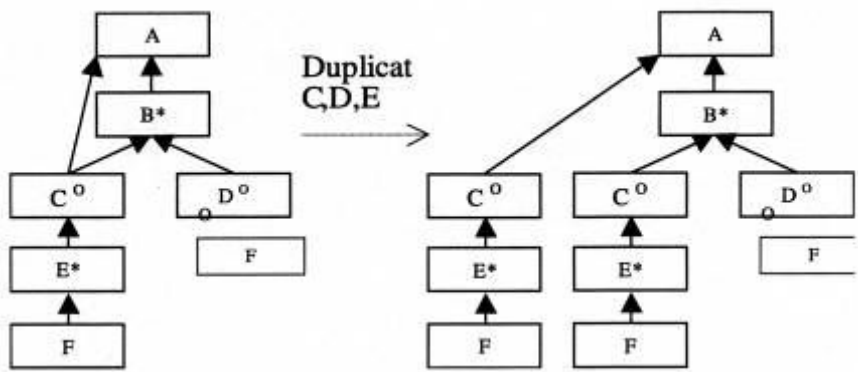

**Figure 5: Chart in the form of a graph**

for C D and E in two different places - as a constituent of A, and as a constituent of B.

The duplication problem can be overcome simply by defining the structure C, D, E, as a separate module, which will occur once in the spreadsheet model. The chart of figure 5 now takes the form of 2 structured modules.

In general, we can always reduce a chart to tree structure by this method, which conveniently produces a unique modularisation of the spreadsheet module, each individual module being expressible in indented form. The modularisation itself, and the relationships between the modules can give a useful overview of complex modules. Figure 6 shows part of an example modular overview for a re-engineering of a normal spreadsheet model. The labels attached to the modules were added after the re-engineering.

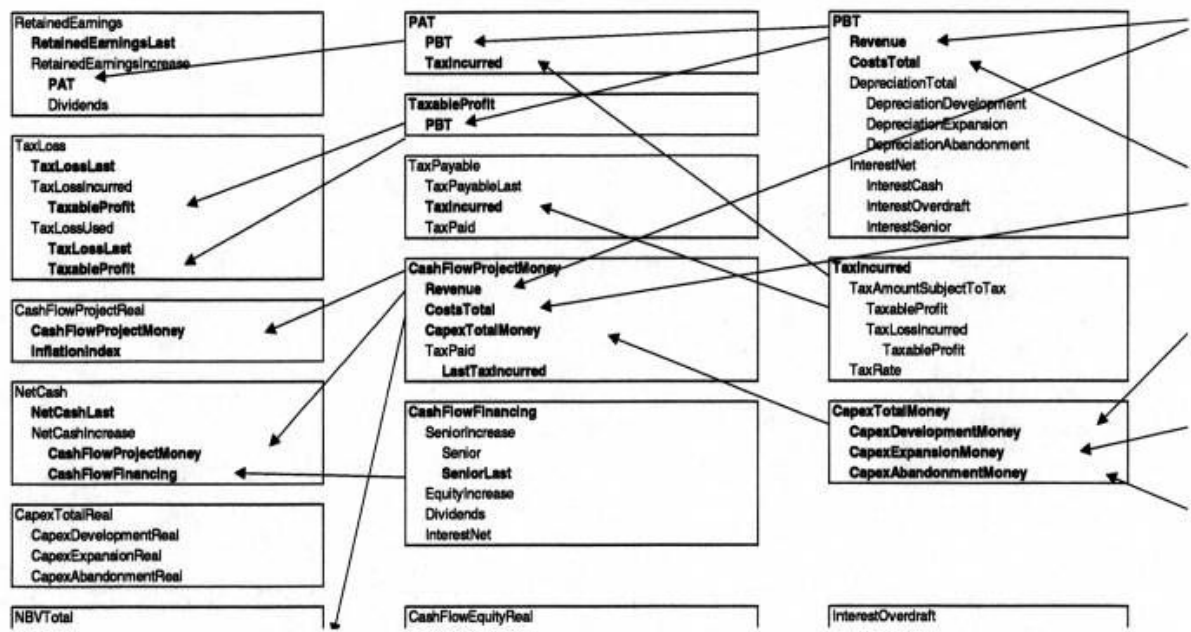

**Figure 6 Part of a Module level overview of the model**

**4. DATA INPUT MODULES**

Data input represents a special problem in spreadsheet design, with its own special requirements. There are reasons why cells for data input should be grouped together in data input modules, separate from the structured modules described above. One reason is to do with the utmost importance of obtaining accurate data entry. The design of this part of the user interface should be as free from constraints as possible; so as not to hinder the main objective: ease of use and absence of data errors. A second reason is that input cells are often referred to by more than one calculated cell. In this case, according to the discussion of the previous section, they should each have the status of a module.

We are however, quite at liberty to put all data input cells into unstructured modules, since there are never any dependencies between them. Any dependency relationship in spreadsheet involves a calculated cell, and either other calculated cells or data input cells. However, they do not exist between data input cells and data input cells. If we do this we end up with the architecture exhibited in figure 7.

The structured spreadsheet modules represent the calculation and display modules. They are the interface accessible with read/write access to the model builder and maintainer, and with read access for the user and auditor. The data entry modules are accessible to the builder, maintainer, auditor with test authorisation, and user with data entry authorisation.

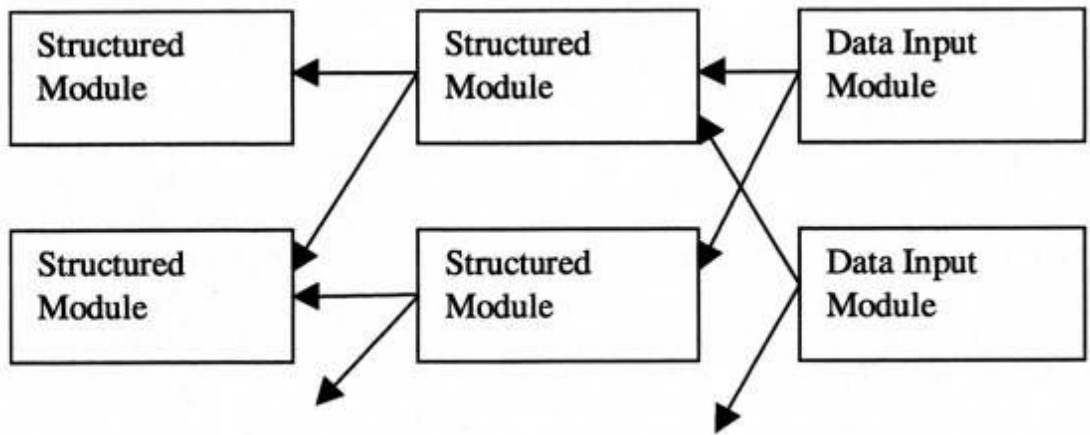

**Figure 7 Architecture of a structured spreadsheet**

## 5. CONCLUSIONS

This paper has described progress on a research project to investigate the use of structured techniques in spreadsheets. It has concentrated on an outline of the main theoretical results obtained, and has indicated their possible use in the construction of sound spreadsheet models. The main results are that structured techniques based on Jackson diagrams may be used with advantage to produce well structured spreadsheets. The techniques give rise to a modularisation principle allowing a decomposition of spreadsheets. The paper shows how individual modules can be structured to advantage, and how an overview of module interactions can be visualised.

The paper has presented an outline only, and has not entered into a discussion of related problems, such as recursive dependency relationships, and practical problems of frequent addition and deletions. We intend to publish a discussion of these problems in a follow up article.

Future work on this project is envisaged on two issues. The first is an investigation of the potential of the structured form for improving the quality of spreadsheets software. The second is work towards an automatic re-engineering tool which can extract information on structure from existing spreadsheets, and translate models into structured form.

**REFERENCES**

1. Warnier, J.D., *Logical Construction of Systems,* Van Nostrand Reinhold, 1981

2. Orr, K. T., *Structured Requirements Definition,* Ken Orr & Associates, 1981

3. Jackson, M., *Principles of Program design,* Academic Press 1975

4. Chen,P., *The entity relationship model - toward a unified view of data. ACM* Trans. Data Base Systems, vol. 1, no. 1, March 1976, pp. 6-36

5. Dahl,O., Dijkstra, E., Hoare, C., *Structured Programming,* Academic Press, 1972

6. Jackson, M., *Principles of Program design,* Academic Press 1975

7. Parnas,D.L., *On the criteria to be used in Decomposing Systems into modules,* CACM, vol. 14, no. 1, April 1972. pp. 221-227.

8. Stevens,W, Myers, G., Constantine,L., *Structured Design,* IBM Systems Journal, vol. 13, No. 2, 1974, pp. 115-139.

9. KPMG Financial Modelling Department (London). Executive Summary: Financial Model Review Survey. KPMG Management Consulting, London, 1997

10. Chadwick, D., "Stop the Subversive Spreadsheet", Internal Auditing and Business Risk magazine, Institute of Internal Auditors United Kingdom, May 2000,

11. Wood, F. Business Accounting 1 (7th Edition). Pitman Publishing, 1996